Distinct Effects of Cr Bulk Doping and Surface Deposition on the Chemical Environment and Electronic Structure of the Topological Insulator $Bi_2Se_3$

Turgut Yilmaz[a,*], William Hines [a], Fu-Chang Sun[b], Ivo Pletikosić[d], Joseph Budnick [a], Tonica Valla[c], and Boris Sinkovic [a]

[a]Department of Physics, University of Connecticut, Storrs, Connecticut 06269, USA

[b]Department of Materials Science and Engineering, University of Connecticut, Storrs, Connecticut 06269, USA

[c]Condensed Matter Physics and Materials Science Department, Brookhaven National Laboratory, Upton, New York 11973, USA

[d]Department of Physics, Princeton University, Princeton, New Jersey 08544, USA

ABSTRACT

In this report, it is shown that Cr doped into the bulk and Cr deposited on the surface of $Bi_2Se_3$ films produced by molecular beam epitaxy (MBE) have strikingly different effects on both the electronic structure and chemical environment. Angle resolved photoemission spectroscopy (ARPES) shows that Cr doped into the bulk opens a surface state energy gap which can be seen at room temperature; much higher than the measured ferromagnetic transition temperature of ≈ 10 K. On the other hand, similar ARPES measurements show that the surface states remain gapless down to 15 K for films with Cr surface deposition. In addition, core-level photoemission spectroscopy of the Bi 5d, Se 3d, and Cr 3p core levels show distinct differences in the chemical environment for the two methods of Cr introduction. Surface deposition of Cr results in the formation of shoulders on the lower binding energy side for the Bi 5d peaks and two distinct Cr 3p peaks indicative of two Cr sites. These striking differences suggests an interesting possibility that better control of doping at only near surface region may offer a path to quantum anomalous Hall states at higher temperatures than reported in the literature.

*yilmaz@phys.uconn.edu

1. INTRODUCTION

Topological insulators (TIs) have attracted tremendous attention recently as a new quantum state of matter. Topological insulators have a band gap and therefore insulating states in the bulk, but possess very special conducting states on the surfaces. These surface states exist and are protected from impurity scattering due to a combination of spin-orbit coupling (SOC) and time-reversal symmetry (TRS). On the surfaces of TIs, inversion symmetry and TRS provide the Kramers degeneracy at momenta points forming gapless surface states characterized by a Dirac-like dispersion relation [1,2]. In addition, this unique electronic structure yields surface states with helical spin texture in which the electron spin is locked perpendicular to the momentum and, therefore, backscattering which requires a spin-flipping process is strongly suppressed. Among the many three dimensional (3D) TIs, $Bi_2Se_3$ is the most prominent candidate for room temperature spintronics applications due to its large bulk band gap ($\approx$ 300 meV) and single Dirac cone at the $\Gamma$(000) point [3].

It is commonly believed that the surface states of TIs are topologically protected against non-magnetic perturbations [2]. On the other hand, for two dimensional (2D) TIs (films), a magnetic field aligned out of the plane breaks the TRS and opens a surface state energy gap at the Dirac point (DP) by lifting the Kramer's degeneracy [4-6]. This was first observed on the 2D TI HgTe by transport measurements [7]. Alternatively, an internal magnetic field can be produced by magnetically doping TIs. Such magnetic TIs represent promising systems for experimentally observing exotic phenomena such as the quantized anomalous Hall effect (QAHE) and giant-magneto-optical effects [8, 9].

Ferromagnetism can be introduced in TI films by three different experimental approaches; doping magnetic impurities into the interior (bulk) during growth, through the surface deposition of magnetic impurities after the growth, or with a magnetic proximity effect [10]. The impact of various magnetic atoms (impurities) on the electronic structure of TIs has been intensively investigated by angle resolved photoemission spectroscopy (ARPES) [5,11-14]. In spite of these various studies, a complete understanding remains elusive. The most puzzling observation made on some bulk doped TIs is the observation of a surface state energy gap not related to a ferromagnetic state [15,16]. For example, a recent study on Mn bulk doped $Bi_2Se_3$ ($Bi_{2-x}Mn_xSe_3$) showed the existence of a surface state energy gap at room temperature, while the magnetic measurements indicated ferromagnetic transition temperatures no greater than ≈ 10 K [15]. In that study, the surface state energy gap was attributed to strong resonant scattering. The Mn 3d orbitals lie in the vicinity of the DP, which results in the localized impurity induced resonance states at and around the DP leading to the opening of a gap.

It appears that opening a surface state energy gap using the surface deposition of magnetic impurities is more challenging than with bulk doping [13,14]. In principle, ferromagnetism can be established through surface deposition by the Ruderman-Kittel-Kasuya-Yosida (RKKY) interaction if the DP is close to the Fermi level [5,6]. Ferromagnetism as a consequence of the RKKY interaction is expected to be stronger than bulk ferromagnetism in TIs because it is inversely proportional to the energy gap [5]. However, ARPES measurements so far have shown that the deposition of Cr on the surface of $Bi_2Se_3$ single crystals does not open a surface state energy gap unlike bulk doped $Bi_2Se_3$ [17]. The impact of Fe surface deposition on the electronic structure of $Bi_2Se_3$ is still under debate [12,13].

In order to further the understanding of the interaction between surface states and magnetic impurities, a comparative study was carried out here to determine the effects of Cr bulk doping and Cr surface deposition on the prototypical TI $Bi_2Se_3$. Films were prepared by molecular beam epitaxy (MBE) and investigated using a combination of ARPES, core-level photoemission spectroscopy, and magnetization. The results from ARPES revealed that Cr bulk doping opens a surface state energy gap which is present even at room temperature, while the measured ferromagnetic transition temperature is ≈ 10 K for both in-plane and out-of-plane orientation. On the other hand, contrary to the case with bulk doping and theoretical expectations, the surface states remain gapless down to 15 K for surface deposited Cr. Photoemission spectroscopy of the Bi 5d, Se 3d, and Cr 3p core levels reveals a different chemical environment for the Cr doped and Cr deposited films.

In addition, the films were characterized using x-ray diffraction (XRD), low energy electron diffraction (LEED), and energy dispersive x-ray spectroscopy (EDXS). The film growth and photoemission studies were carried out at the U5 beam line of the National Synchrontron Light Source (NSLS-I4), Brookhaven National Laboratory (BNL), while the magnetization studies were carried out in the Physics Department at the University of Connecticut (UConn).

## 2. EXPERIMENTAL APPARATUS AND PROCEDURE

### 2.1. Sample Preparation

High quality $Bi_2Se_3$ films were grown on $Al_2O_3$ (0001) substrates using MBE in which high purity Bi (99.9999%) and Se (99.9999%) were co-evaporated from resistively heated crucibles. A two step growth method was employed in which 2 quintuple layers (QL) of $Bi_2Se_3$ were initially grown at 117 ºC, followed by the growth of the remaining material at 192 ºC [18]. After the growth, the films were kept at 192 ºC for an additional 10 minutes in order to promote surface quality. Surface deposited films were produced by depositing various thicknesses of Cr (up to 1.2 monolayer (ML)) on the surface of $Bi_2Se_3$ films at room temperature by using an electron-beam evaporator. The Cr bulk doped samples were prepared by introducing the Cr atoms into the interior during the growth of the film. Evaporation rates from the sources were calibrated with a quartz microbalance sensor. The stoichiometry of the bulk doped samples was measured by EDXS and found to be $Bi_{1.95}Cr_{0.05}Se_3$ and $Bi_{1.84}Cr_{0.16}Se_3$. In this work, the MBE growth chamber had a base pressure $2 \times 10^{-11}$ torr; prior to growth, the $Al_2O_3$ (0001) substrates were annealed at 650 ºC for 2 hours and then annealed at 850 ºC for another 30 minutes to produce an atomically flat surface, free of contamination. The ultra-high vacuum MBE growth chamber and the photoemission chamber were vacuum interlocked in order to avoid surface contamination. Therefore, all photoemission measurements were performed on effectively *in-situ* grown samples. Films that were removed from the chambers for the SQUID magnetometry, XRD, LEED, and EDXS measurements, were capped with 2 nm to 10 nm of Se.

2.2. Photoemission Apparatus

Angular resolved photoemission spectroscopy and core-level photoemission spectroscopy were carried out with 52 eV and 104 eV synchrotron radiations, respectively. The photoemission chamber consists of a Scienta SES-2002 electron analyzer and, with the use of synchrotron radiation, has 10 meV energy and 0.15º angular resolution. Binding energies of core levels were calibrated relative to the Fermi level with ± 0.02 eV error throughout. The Bi 5d and Se 3d peaks were fit with a linear combination of Lorentzian and Gaussian line shapes, while a Doniach-Sunjic line shape was used for the Cr 3p peaks. A Shirley type of background was applied to all core-level spectra.

2.3. Magnetic Characterization

Measurements of the dc magnetization were carried out for magnetic fields $-50$ kOe $\leq$ H $\leq$ $+50$ kOe over the temperature range 5.0 K $\leq$ T $\leq$ 300 K using a superconducting quantum interference device (SQUID) MPMS-5 magnetometer from Quantum Design. Measurements of the magnetization temperature dependence and hysteresis loops were obtained with the magnetic field applied both in the plane (parallel to the plane) and out of the plane (perpendicular to the plane) of the film. The $Al_2O_3$ (0001) substrates (5 mm × 5 mm × 0.5 mm) were mounted on a quartz rod for the parallel measurements and in a straw for the perpendicular measurements. In order to obtain a background calibration, corresponding magnetic measurements were also made on a clean $Al_2O_3$ (0001) substrate with just a 2 nm Se cap. As is the case with this work, great care must be taken to avoid contamination when making measurements involving nanoscale magnetism (e.g., moment values $< 10^{-4}$ emu) [19]. Furthermore, the SQUID response curves are somewhat distorted as the samples are in the form of finite planes and not ideal dipoles [20]. Finally, magnetic artifacts can occur if the sample chamber contains even small amounts of residual oxygen [21]. These issues, all of which are important for the work reported here, are discussed in detail in Refs. [19-21].

3.      RESULTS AND ANALYSIS

3.1.    Angle Resolved Photoemission Spectroscopy

Fig. 1 shows the electronic structure obtained along the Γ−M direction at room temperature by ARPES for a 23 QL thick pristine $Bi_2Se_3$ film and two 23 QL thick Cr doped $Bi_{2-x}Cr_xSe_3$ films with x = 0.05 and 0.16. A 52 eV photon energy was used to obtain the ARPES spectra in order to enhance the spectral strength of the surface states and mostly eliminating the conduction band contribution [22,23]. In Fig. 1(a), the unique energy spectrum for the pristine $Bi_2Se_3$ film shows a DP located at Γ(000) with a corresponding binding energy of 290 meV. The gapless and linear surface states marked with the solid blue lines in Fig. 1(a) indicate the Dirac-like dispersion for pristine $Bi_2Se_3$. Cr bulk doping has an immediate impact on the electronic structure of $Bi_2Se_3$. The most notable observation in the ARPES spectra is that a relatively small amount (x = 0.05) of Cr doping into the bulk of $Bi_2Se_3$ destroys the Dirac-like dispersion and opens a surface state energy gap indicated by the solid black lines in Figs. 1(b) and 1(c). The size of the energy gap increases with the Cr content and is measured to be ≈ 50 meV and ≈ 100 meV for x = 0.05 and 0.16, respectively. Furthermore, Cr bulk doping induces a pronounced momentum broadening of the surface states which also increases with the Cr content. It can be seen in Fig. 1(c) that the broadening results in considerable overlapping between the +**k** and −**k** energy states for x = 0.16 which compromises their clear identification.

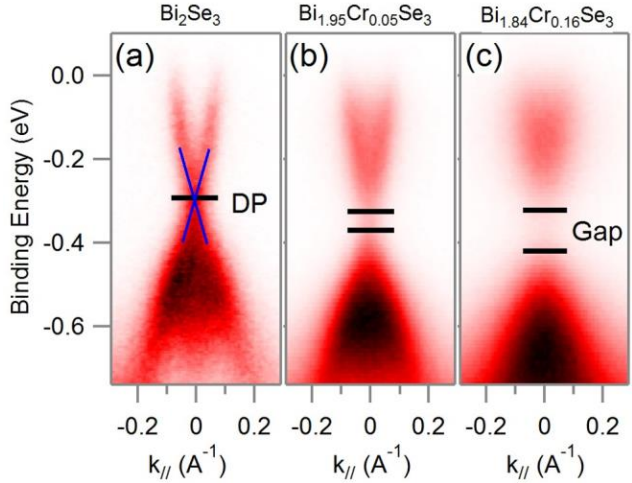

Fig. 1. (Color Online) Electronic structure of 23 QL thick Bi2−xCrxSe3 films taken along the Γ−M direction at room temperature using ARPES with 52 eV radiation: (a) pristine (x = 0), (b) x = 0.05, and (c) x = 0.016. In (a), the solid blue lines represent the surface states and the Dirac point is shown by a single black line. The surface state energy gap is indicated by the solid black lines.

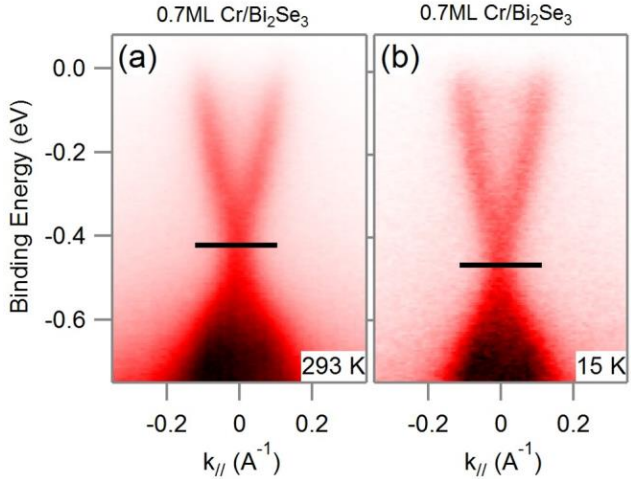

Fig. 2. (Color Online) Electronic structure of a 23 QL thick $Bi_2Se_3$ film covered with 0.7 ML Cr taken along the Γ−M direction using ARPES with 52 eV radiation: (a) room temperature and (b) 15 K. Solid black lines represent the Dirac Point.

In order to compare the effect of Cr surface deposition with that of Cr bulk doping described above, ARPES spectra were obtained from a 23 QL thick $Bi_2Se_3$ film covered with 0.7 monolayer (ML) of Cr. The ARPES spectra were again measured with 52 eV radiation, both at room temperature and 15 K. As shown in Fig. 2(a), the deposition of 0.7 ML of Cr onto the surface of $Bi_2Se_3$ shifts the DP to ≈ 430 meV, an ≈ 140 meV increase in the binding energy compared to pristine $Bi_2Se_3$. However, unlike the case of Cr bulk doping, the room temperature spectrum does not show an energy gap opening for Cr surface deposition. In order to test for the existence of a low ferromagnetic ordering temperature, and consequently the opening of an energy gap, an ARPES spectrum was measured also at 15 K (see Fig. 2(b)). Although the DP has an additional increase in binding energy of ≈ 40 meV (to ≈ 470 meV), there is still no appearance of an energy gap opening at the DP. In summary, the ARPES results indicate robust gapless surface states for $Bi_2Se_3$ upon the surface deposition of Cr.

3.2. Core-Level Photoemission Spectroscopy

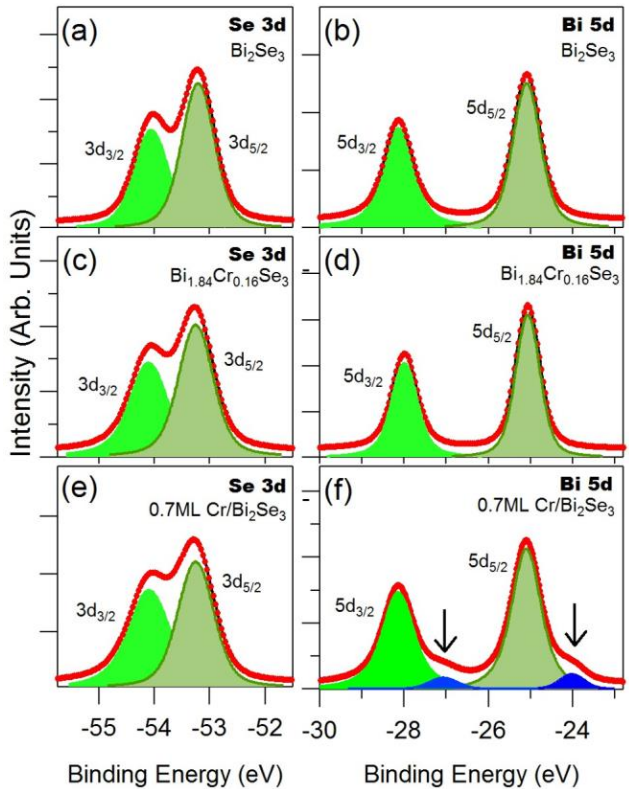

Fig. 3. (Color Online) Se 3d and Bi 5d core-level photoemission spectra obtained from 23 QL thick films: (a) and (b) pristine $Bi_2Se_3$; (c) and (d) Cr bulk doped $Bi_{1.84}Cr_{0.16}Se_3$; (e) and (f) 0.7 ML Cr surface deposited $Bi_2Se_3$. The spectra were obtained with a 104 eV photon energy at room temperature. The Se 3d and Bi 5d peaks (blue filled areas) were fit with a linear combination of Lorentzian and Gaussian line shapes (superposition

indicated by dotted red line). A Shirley type of background was applied to all core-level spectra. Black arrows indicate shoulder-like structures on the Bi 5d peaks.

The elemental composition, elemental oxidation state, and local electronic structure in the near-surface region (≈ 1 nm) of the films was studied by core-level photoemission spectroscopy using 104 eV photons. As was the case for the electronic structure revealed in the ARPES spectra described above, Cr bulk doping and Cr surface deposition can result in different effects on the local, chemical environment in $Bi_2Se_3$ (see Fig. 3). The Se 3d and Bi 5d core-level spectra obtained from pristine $Bi_2Se_3$ are shown in Figs. 3(a) and 3(b), respectively. The spin-orbit split Bi $5d_{5/2}$ and $5d_{3/2}$ peaks have binding energies of 25.0 eV and 28.1 eV, respectively, while the binding energies of the Se $3d_{5/2}$ and $3d_{3/2}$ peaks are measured to be 53.2 eV and 54.1 eV, respectively. The Cr bulk doped $Bi_{1.84}Cr_{0.16}Se_3$ sample shows a chemical environment very similar to that for the pristine $Bi_2Se_3$ sample except that the Se 3d (see Fig. 3(c)) and Bi 5d (see Fig. 3(d)) core levels are shifted to higher binding energies by ≈ 80 meV. However, there was no sign of the formation of chemically different peaks or significant broadening (full width at half maximum) of the Se 3d or Bi 5d peaks. This suggests the absence of intercalated or Cr-substituted Se. This is also supported by XRD results in which it is observed that the c-axis lattice parameter remains essentially unchanged as the Cr content is increased.

Figures 3(e) and 3(f) show the Se 3d and Bi 5d core-level photoemission spectra obtained from a 0.7 ML Cr surface deposited $Bi_2Se_3$ film, respectively. As was the case for the Cr bulk doped films, the Se 3d and Bi 5d peaks for the Cr surface deposited films are shifted to higher binding energy compared to those for pristine Bi$_2$Se$_3$. However, in contrast to the core-level photoemission spectra reported above for the pristine and Cr doped films, Cr surface deposition results in the appearance of a shoulder structure on the low binding energy sides of the main Bi 5d$_{5/2}$ and 5d$_{3/2}$ peaks (indicated by the black arrows and blue-filled areas in Fig. 3(f)). These new structures (peaks), which are located at 23.9 eV and 27.0 eV, are comparable to the binding energies of Bi metal. Similar Bi 5d shoulder peaks were reported in the core-level spectra for Fe surface deposited Bi$_2$Se$_3$ [13] and Fe surface deposited Bi$_2$Te$_3$ [24]. In the former, the shoulder peaks were attributed to Bi with a lower oxidation state than 3+ [13]. In the latter, the shoulder peaks were attributed to either Bi metal formation or Bi sub-tellurides [24]. As discussed below, the formation of Bi metal here with the surface deposition of Cr on $Bi_2Se_3$ is one possibility. In order to explore the existence of more than one Cr site for the surface deposited Cr, Cr 3p core-level photoemission spectra were obtained from the films with various Cr layer thickness (see Fig. 4). It can be seen that two distinct Cr 3p peaks appear (at ≈ 40.4 eV and ≈ 41.9 eV) which are attributed to two distinct chemical environments and not simply a spin-orbit splitting (the Cr 3p spin-orbit splitting is too small). The intensity of the two peaks increases with the Cr thickness. As discussed below, the two Cr 3p peaks are attributed to two distinct Cr sites, with the possibility of one of the sites being in the bulk.

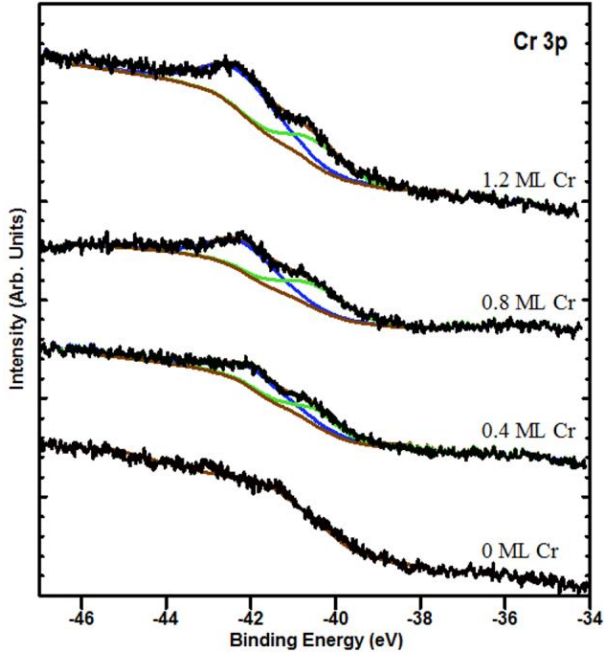

Fig. 4. (Color Online) Cr 3p core-level photoemission spectra obtained from 23 QL thick Bi$_2$Se$_3$ films with indicated thickness of surface deposited Cr. The spectra were obtained with a 104 eV photon energy at room temperature. The two distinct Cr 3p peaks were fit with a Doniach-Sinjoc line shape (blue and green lines). A Shirley type of background was applied to all core-level spectra (brown lines).

3.3. Magnetization

As described above, ARPES results from the Cr bulk doped films $Bi_{1.95}Cr_{0.05}Se_3$ and $Bi_{1.84}Cr_{0.16}Se_3$ showed the existence of a surface state energy gap at room temperature. In order to explore the magnetic behavior and look for possible long-range ferromagnetic order, detailed magnetization measurements were carried out on 46 QL thick pristine $Bi_2Se_3$ and Cr doped $Bi_{1.84}Cr_{0.16}Se_3$ films. The films were again prepared on $Al_2O_3$ (0001) substrates and capped with 2 nm of Se before removal from the MBE chamber. Figure 5(a) shows the magnetic field dependence of the total magnetic moment −10 kOe ≤ H ≤ + 10 kOe at temperatures T = 300 K, 100 K, and 5.0 K for the pristine $Bi_2Se_3$ film. The magnetic field was applied parallel to the film surface. It can be seen that, for the three temperatures, the magnetic behavior is simply linear through the origin with neither hysteresis nor any sign of a ferromagnetic component. The magnetic behavior is dominated by the diamagnetic response of the $Al_2O_3$ (0001) substrate. The slopes of the curves become slightly more negative as the temperature is increased indicating trace amounts of paramagnetic impurities. In addition, an $Al_2O_3$ (0001) substrate with only the 2 nm Se cap was measured at the same three temperatures and yielded similar results.

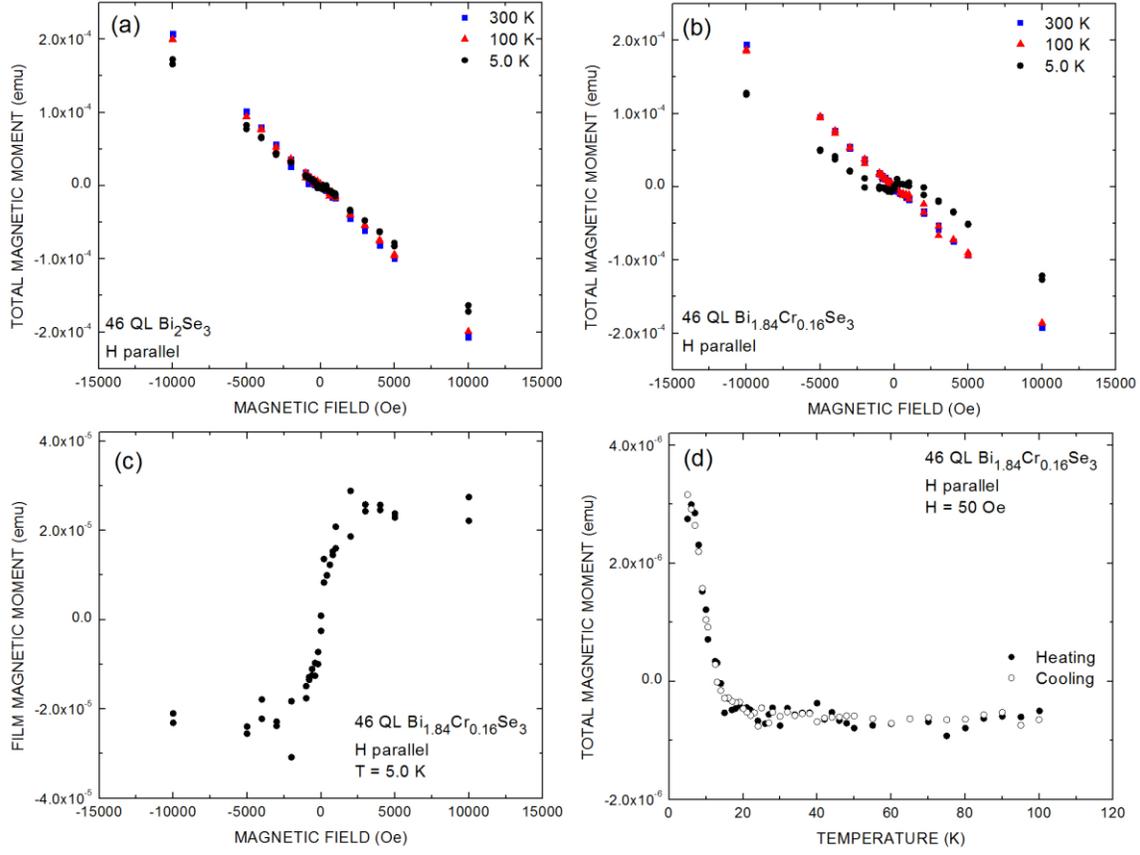

Fig. 5. (Color Online) Total magnetic moment (emu) versus magnetic field (Oe), magnetic hysteresis loops, for a 46 QL thick (a) $Bi_2Se_3$ film and (b) $Bi_{1.84}Cr_{0.16}Se_3$ film on an $Al_2O_3$ (0001) substate: solid blue squares – 300 K, solid red triangles – 100 K, and solid black circles – 5.0 K. The magnetic field is applied in the plane of the film. Only the curve for the $Bi_{1.84}Cr_{0.16}Se_3$ film at 5.0 K shows a small ferromagnetic component. (c) Film only magnetic moment (emu) versus magnetic field (Oe), magnetic hysteresis loop, for the $Bi_{1.84}Cr_{0.16}Se_3$ film at 5.0 K which is obtained by subtracting the diamagnetic background from the T = 5.0 K curve in Fig. 5(b). (d) Zero-field-cooled/field-cooled curves showing the total magnetic moment (emu) versus temperature (K) for the $Bi_{1.84}Cr_{0.16}Se_3$ film in a magnetic field of 50 Oe applied in the plane of the film: solid circles – heating and open circles – cooling. A magnetic transition is indicated at ≈ 10 K.

Figure 5(b) shows the magnetic field dependence of the total magnetic moment $-10$ kOe $\leq$ H $\leq$ $+$ 10 kOe at temperatures T = 300 K, 100 K, and 5.0 K for the $Bi_{1.86}Cr_{0.16}Se_3$ film. The magnetic field was applied parallel to the film surface. It can be seen that, for T = 300 K and 100 K, the magnetic behavior is again linear through the origin; however, a small, but clear, ferromagnetic component appears for T = 5.0 K. Figure 5(c) shows the magnetic field dependence of just the film moment at T = 5.0 K which was obtained by subtracting the diamagnetic component from the total moment curve in Fig. 5(b). The ferromagnetic component appears "soft", i.e., no significant coercive field or remnant magnetization, with a saturation magnetic moment $2.4 \times 10^{-5}$ emu. From the volume of the film ($1.1 \times 10^{-6}$ cm$^3$) and the mass density of $Bi_2Se_3$ (7.51 g/cm$^3$), a value of 2.1 $\mu_B$ is calculated for the Cr magnetic moment. As pointed out by W. Liu, et al. [25], various magnetic studies on epitaxial, Cr-doped $Bi_2Se_3$ report a Cr magnetic moment $\approx$ 2 $\mu_B$ which is less than the Hund's rule value of 3.0 $\mu_B$ for $Cr^{3+}$ which is substituted into the Bi sites.

Figure 5(d) shows the zero-field-cooled/field-cooled (ZFC/FC) magnetization curves obtained with a magnetic field H = 50 Oe applied parallel to the film surface for the $Bi_{1.86}Cr_{0.16}Se_3$ film. The ZFC/FC protocol used here involves cooling the sample from high temperature ($\geq$ 100 K) to 5.0 K in zero field and then applying a magnetic field of 50 Oe. The ZFC curve measurement is obtained during heating to 100 K, followed by cooling back to 5.0 K (FC curve measurement). The magnetic response is completely reversible and indicates a magnetic transition $T_C \approx$ 10 K.

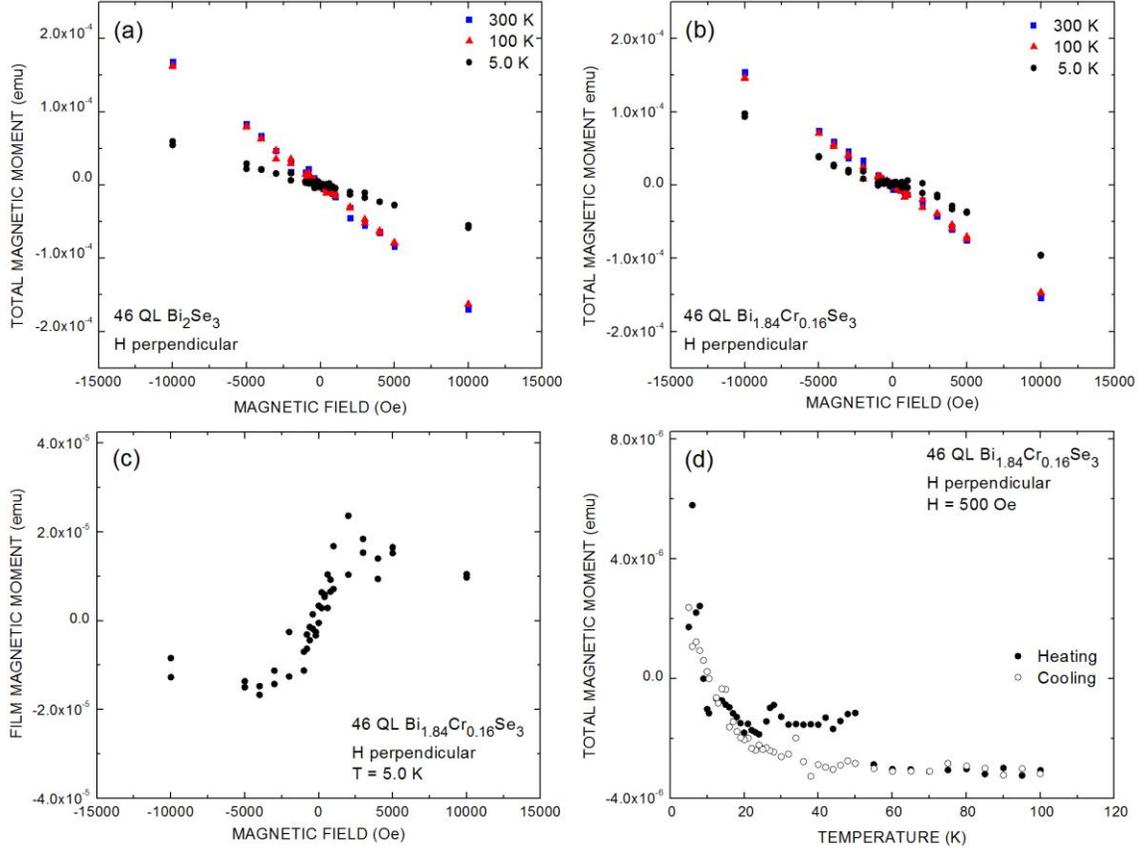

Fig. 6. (Color Online) Total magnetic moment (emu) versus magnetic field (Oe), magnetic hysteresis loops, for a 46 QL thick (a) $Bi_2Se_3$ film and (b) $Bi_{1.84}Cr_{0.16}Se_3$ film on an $Al_2O_3$ (0001) substrate: solid blue squares – 300 K, solid red triangles – 100 K, and solid black circles – 5.0 K. The magnetic field is applied perpendicular to the plane of the film. Only the curve for the $Bi_{1.84}Cr_{0.16}Se_3$ film at 5.0 K shows a small ferromagnetic component. (c) Film only magnetic moment (emu) versus magnetic field (Oe), magnetic hysteresis loop, for the $Bi_{1.84}Cr_{0.16}Se_3$ film at 5.0 K which is obtained by subtracting the diamagnetic background from the T = 5.0 K curve in Fig. 6(b). (d) Zero-field-cooled/field-cooled curves showing the total magnetic moment (emu) versus temperature (K) for the $Bi_{1.84}Cr_{0.16}Se_3$ film in a magnetic field of 500 Oe applied in the plane of the film: solid circles – heating and open circles – cooling. A magnetic transition is indicated at ≈ 10 K.

Finally, similar results were obtained for the $Bi_2Se_3$ and $Bi_{1.86}Cr_{0.16}Se_3$ films on $Al_2O_3$ (0001) substrates when the magnetic field was applied <u>perpendicular</u> to the film surface (see Figs. 6(a), 6(b), 6(c), and 6(d)). It was found that making magnetic measurements when the applied magnetic field was perpendicular to the film surface, and hence the $Al_2O_3$ (0001) substrate, was more problematic than with the parallel orientation. The value obtained for the Cr moment was smaller for the perpendicular orientation, i.e., 1.3 $\mu_B$. A possible explanation is the existence of a demagnetization effect which would not be present with the parallel orientation. As pointed out above, the perpendicular orientation produced some distortion in the SQUID response curve [20]. Also, as pointed out above, great care must be taken when making measurements involving nanoscale magnetism such as the case for thin films [19-21]. In spite of these challenges, the magnetic behavior for the perpendicular and parallel orientations appears to be similar.

4. DISCUSSION

A predominant idea of how to break the TRS appearing in the literature to date is to modify the topological surface states by introducing 3d transition metal elements (e.g., V, Cr, Mn, or Fe) into the bulk (interior) or on the surface of TI films [5,12]. However, in spite of several recent experimental studies, principally ARPES, the introduction of a gap in the surface states through doping/deposition of magnetic impurities remains controversial, a gap has been observed in some studies but not in others. Furthermore, the effect of magnetic versus non-magnetic surface impurities is not clear. In an attempt to reconcile the apparently inconsistent experimental results, Black-Schaffer et al. [26] have investigated the effect on the surface states from both magnetic and potential scattering.

In this report, the different effects of Cr bulk doping and Cr surface deposition on the electronic structure and chemical environment of the TI $Bi_2Se_3$ have been investigated in order to further the understanding of the interaction between the surface states and magnetic impurities. The combined study of photoemission and magnetization revealed that Cr bulk doping opens a surface state energy gap which can be observed even at room temperature, while the measured ferromagnetic transition temperature is ≈ 10 K for both in-plane and out-of-plane orientation. On the other hand, in contrast to Cr bulk doping as well as theoretical expectations, for Cr surface deposition the surface states remain gapless down to 15 K.

The most striking result from the ARPES measurements on the Cr-doped films is the observation of a gap at both 100 K and room temperature. This is far above the transition temperature for long-range ferromagnetic order observed here ($T_C$ ≈ 10 K) or in the literature (highest $T_C$ = 35 K [27]). This result is in contradiction with previous theoretical work which has suggested that a long-range ferromagnetic order with a magnetic moment out-of-the-plane of the film is required to break the TRS and open an energy gap at the Dirac point [5,12]. One possible explanation is that the replacement of Bi atoms with much lighter Cr atoms results in a quantum phase transition (QPT) from a topologically non-trivial insulator to a trivial insulator and the observed gap at the DP is simply the bulk gap of an ordinary insulator [28]. Such quantum phase transition (QPT) from a topologically non-trivial insulator to a trivial insulator has been found in $(Bi_{1-x}In_x)_2Se_3$ and $(Bi_{1-x}Sb_x)_2Se_3$ by tuning the SOC strength with the In and Sb concentration [29]. The other possible explanation for the opening of a gap at the DP above the magnetic ordering temperature is that impurity resonances destroy the topological protection of the surface states [26]. Such ideas were experimentally examined in a very recent study of Mn-doped TI films by Sánchez-Barriga et al. [15]. They used a combination of ARPES and SQUID magnetometry to study the electronic band structure and magnetic properties of MBE-grown $(Bi_{1-x}Mn_x)_2Se_3$ ($0 \leq x \leq 0.08$) films. They also find that with Mn doping, a large band gap ($\approx$ 200 meV) opens at the DP point even at 300 K, which is far above their measured $T_C \approx$ 10 K. Along with *ab initio* calculations, they conclude that it is the impurity-induced resonance states remove the Dirac point of the surface states and not long-range ferromagnetic order. Furthermore, they found no indication for a reversal of band inversion due to the Mn doping, thereby eliminating the possibility of the gap being an ordinary insulating gap. Finally, they report that even low concentrations of nonmagnetic In opens a surface band gap.

Although Cr is one of the most used dopant in the investigation between the surface states of TIs in general, and $Bi_2Se_3$ in particular [16,17,25,27,30-35], core-level photoemission spectroscopy on both Cr bulk doped and Cr surface deposited $Bi_2Se_3$ has not been reported to date. Here, Bi 5d, Se 3d, and Cr 3p core-level photoemission shows a clear difference in the chemical environment for these two cases. In the case of Cr bulk doping, the decrease in the intensity of the Bi 5d peaks with increasing Cr content supports the idea that Cr occupies a single site and most likely substitutes for Bi [Ref. 35 and references therein]. However, in the case of Cr surface deposition, a new shoulder-like peak appears on the lower binding side for each of the two main Bi 5d peaks; the intensity of these satellite peaks increases with the Cr content. As mentioned above, the two new Bi 5d satellite peaks have 23.9 eV and 27.0 eV binding energies, which are comparable to the Bi metal 5d peaks. Furthermore, two distinct Cr 3p peaks appear whose splitting is too large to be a result of spin-orbit coupling. The Bi 5d and Cr 3p core-level photoemission spectra for surface deposited Cr clearly indicate the existence of two distinct Cr sites. One possibility interpretation of the Bi 5d shoulders are that they are consequence of Bi metal formation resulting from the substitution of Cr for Bi in the bulk just below the surface layer. There exists in the literature some theoretical work which suggests that Bi-site substitution is energetically favorable for Cr [36]. However, the ARPES results do not show the broadening of the surface states or opening of a surface state energy gap as is the case for Cr bulk doping. Another possibility with Cr surface deposition is the existence of two (or more) distinct Cr sites on the surface as proposed by Wang et al. [17]. The top layer of $Bi_2Se_3$ is known to be a Se layer [3,17,37]. Even with Cr on top of the Se layer, a strong interaction between the Cr and Se could weaken the bonding between Se and the Bi atoms below [13]. Therefore, the shoulder structure on the lower binding energy side of the Bi 5d peaks could originate (indirectly) from the formation of new bonds between the Cr and Se atoms since it is unlikely that Cr on the surface can directly alter the Bi 5d core levels to such an extent that shoulder peaks occur at a binding energy ≈ 1 eV below the main peaks.

There are two reports in the literature concerning the surface deposition of magnetic impurities on single crystal samples of $Bi_2Se_3$. Scholz et al. [13] studied the effect of Fe impurities deposited on $Bi_2Se_3$ at room temperature and 8 K using ARPES and core-level photoemission. For both temperatures, they observe that the surface states remain intact and gapless. They also observe shoulders on the Bi 5d peaks which depend on the deposition temperature and are a consequence of different Fe environments such as Fe islands or subsurface sites. Wang et al. [17] combine APRES experiments on a freshly cleaved $Bi_2Se_3$ single crystal surface after Cr deposition along with density functional theory calculations. They show that the Cr surface deposition does not open a gap and conclude that the origin is attributed to different Cr occupation sites from the bulk Cr doped samples. Our results on Cr surface deposition on $Bi_2Se_3$ films are consistent with these reports from $Bi_2Se_3$ crystals and furthermore show the complexity of the surface structure that needs to be part of any theoretical model explaining the electronic properties of these systems.

For the case of the Cr doped films reported here, previous results from conventional (SQUID) magnetometry have consistently yielded low ferromagnetic ordering temperatures (≤ 35 K) and Cr magnetic moment values significantly lower than the spin-only 3 $\mu_B$ predicted by Hund's rules for $Cr^{3+}$. Haazen et al. [32] reported the first observation of ferromagnetism in the Cr-doped system $Cr_x(Bi_2Se_3)_{(1-x)}$ with maximum Curie temperature of 20 K for 5.2% Cr, and maximum Cr moment of 2.2 $\mu_B$ for the lowest doped film (1.3% Cr). They attribute the low Cr moment value to not all of the Cr contributing to the ferromagnetism (e.g., the partial formation of some antiferromagnetic phase such as Cr or CrSe), or perhaps, the Cr being in a valence state other than $Cr^{3+}$. Collins-McIntyre et al. [33] observed in-plane ferromagnetic behavior for a $Bi_{1.76}Cr_{0.24}Se_3$ film with $T_C$ = 8.5 K and a Cr moment value of 2.1 $\mu_B$. They also attribute the "missing moment" to the possibility that some of the Cr does not participate in the long-range ferromagnetic order. Figueroa et al. [35] using element-specific x-ray absorption techniques (EXAFS, XAS and XMCD) studied a thick (100 nm) $Bi_{1.76}Cr_{0.24}Se_3$ film and claim that the Cr is principally divalent in the "bulk" and principally trivalent on the surface region (top ~ 5 nm) with bulk and surface moment values of 3.64 $\mu_B/Cr_{bulk}$ and 1.91 $\mu_B/Cr_{surf}$, respectively, yielding an average value of 2.84 $\mu_B/Cr_{ave}$ for the entire film. Using both magneto-transport and SQUID magnetometry, Kou et al. [27] report "unconventional" ferromagnetic order below $T_C$ = 35 K for a $Bi_{1.90}Cr_{0.10}Se_3$ film but no value is given for the Cr magnetic moment. More recently, Liu et al. [25] carried out a detailed XMCD study, along with density functional theory simulations and obtain a total moment 2.00 $\mu_B/Cr$ (spin moment 2.05 $\mu_B/Cr$ and orbital moment -0.05 $\mu_B/Cr$), suggesting that its low value is due to the spontaneous coexistence of ferromagnetic and antiferromagnetic Cr defects in $Bi_{2-x}Cr_xSe_3$. Finally, Choi et al. [34] have studied the transport and magnetic properties of Cr-, Fe-, and Cu- doped $Bi_2Se_3$ bulk single crystals. A Curie-Weiss fit of their high-temperature magnetization data (100 K $\leq$ T $\leq$ 300 K) for $Bi_{1.85}Cr_{0.15}Se_3$, yields a Curie-Weiss temperature $\Theta$ = −229.8 K indicating strong antiferromagnetic interactions; but no observation of an ordering temperature ($T_N$). They reported value for $\mu_{eff}$ = 0.95 $\mu_B$/f.u. which is not consistent with divalent Cr having a moment of 4.9 $\mu_B$ as claimed. Our SQUID magnetometry of bulk doped $Bi_2Se_3$ films discussed above agrees with majority of the reported studies on the magnetic moment being in the ~2 $\mu_B$/Cr range, while variation in that value reported in the literature indicates the sample complexities even in this bulk doped system, which deserves further studies.

5. CONCLUSION

In summary, the inclusion of 3d metal elements (e.g, Cr) into a TI such as $Bi_2Se_3$ by bulk doping or surface deposition results in two very different system with different local chemical environment and electronic structure. The bulk Cr doping results in large gap at DP well above the ferromagnetic transition temperature, indicating that the gap is not due to TRS breaking. On the other hand, the deposition of Cr on the surface of $Bi_2Se_3$ results in no observable DP gap at lowest measuring temperature (15 K), which however may still be above the onset of ferromagnetic state in that system. Furthermore, in the first case the topological protection is completely destroyed while in the later there is no measurable effect of Cr on the topological states. This gives an interesting possibility that better control of Cr doping at only the near surface region may lead to gap opening at DP by break the time-reversal symmetry and emergence of the quantized anomalous Hall state, which has been the major goal in this research field.

ACKNOWLEDGEMENTS

This work was supported by the US Department of Energy, Office of Basic Energy Sciences, Contract No. DE-AC02-98CH10886 and the ARO MURI program, Grant No. W911NF-12-1-0461 and UCONN-REP Grant No. 4626510.